\documentclass[12pt]{article}

\usepackage{tikz}
\usetikzlibrary{arrows}

\usepackage{caption}
\usepackage{wrapfig}
\usepackage{bm}
\usepackage{graphicx}
\usepackage{amssymb,mathrsfs}
\usepackage{amsthm}
\usepackage{amsmath} 
\usepackage{amsmath, amsfonts} 
\usepackage{stmaryrd}
\usepackage{float}
\usepackage{subfig}
\usepackage{chngcntr}

\counterwithout{figure}{section}
\counterwithout{figure}{subsection}

\newcommand{\de}{\partial}
\newcommand{\dst}{\displaystyle}
\newcommand{\ep}{\varepsilon}
\newcommand{\be}{\bm \ep}
\newcommand{\ei}{\ep_{tu}}
\newcommand{\ee}{\ep_{ex}}
\newcommand{\ec}{\ep_{in}}
\newcommand{\alphat}{\widetilde{\alpha}}
\newcommand{\nonu}{\nonumber}
\newcommand{\bt}{\bm \tau}
\newcommand{\E}{\bm E}
\newcommand{\D}{\bm D}
\newcommand{\M}{\bm M}
\newcommand{\I}{\bm I}
\newcommand{\ui}{u_{in}}
\newcommand{\ue}{u_{ex}}
\newcommand{\ut}{u_{tu}}
\newcommand{\rf}[1]{(\ref{#1})} 
\renewcommand{\leq}{\leqslant}
\renewcommand{\geq}{\geqslant}
\newcommand{\la}{\langle}
\newcommand{\ra}{\rangle}
\renewcommand{\P}{P_{m,n}}
\newcommand{\x}{{\bm u}}
\newcommand{\y}{{\bm v}}

\newcommand{\G}{\mathcal{G}}
\newcommand{\C}{\mathbb C}
\newcommand{\re}{\mbox{Re\,}}
\newcommand{\im}{\mbox{Im\,}}
\newtheorem{theorem}{Theorem}

\begin{document}

\title{EFFECTIVE PROPERTIES OF PERIODIC TUBULAR STRUCTURES\thanks{
This work was supported, in part, by funds provided by the University of North Carolina at Charlotte.}} 

\author{Yuri~A. Godin \\
Department of Mathematics and Statistics \\
University of North Carolina at Charlotte \\
Charlotte, NC 28223, USA \\
email: ygodin@uncc.edu}

\maketitle

\begin{abstract}
A method is described to calculate effective tensor properties of a periodic array
of two-phase dielectric tubes embedded in a host matrix. The method uses Weierstrass'
quasiperodic functions for representation of the potential that considerably facilitates
the problem and allows us to find an exact expression for the effective tensor. For weakly 
interacting tubes we obtain Maxwell-like approximation of the effective parameter which is 
in very good agreement with experimental results in considered examples. 
\end{abstract}

%


\pagestyle{myheadings}
\thispagestyle{plain}
\markboth{YURI~A. GODIN}{EFFECTIVE PROPERTIES OF PERIODIC TUBULAR STRUCTURES}

\section{Introduction}

The problem of evaluating the effective properties (permittivity, conductivity, etc.) of periodic 
heterogeneous materials has been extensively investigated. Its solution for noninteracting 
particles was suggested by Maxwell \cite{M:73}, which has become ubiquitous in physics and engineering 
as well as an indispensable benchmark asymptotics. Despite apparent limitations, it provides a good approximation
in a certain range of parameters for the estimation of optical properties of square lattice of carbon nanotubes 
\cite{G:97},\cite{R:05} as well as optical properties of artificially engineered microstructured materials \cite{M:03}.

The seminal paper of Rayleigh \cite{R:92} predestined the development in this area for many decades to come. 
It contained the ideas of the multipole expansion method, relation of the potential with the elliptic functions, 
its application to elasticity and wave propagation. Rayleigh's method was extended to a regular arrays of
cylinders \cite{PMM:79},\cite{Mc:86},\cite{N:93} as well as to the dynamic problems \cite{Z:02}. 

Application of Rayleigh's approach to arbitrary lattices, however, encounters two obstacles. The distribution
of stream lines is not known for the medium whose effective properties are described by a tensor. As a result,
the method used in \cite{R:92} for evaluation of a scalar is not applicable for determination of the effective
 tensor. Next, the method entails conditionally convergent sum whose summation order is obscure. That hampers 
further development of the method.   

The advantages of application of the elliptic and meromorphic functions to the problems of determination of 
the effective properties of perforated plates and shells had been clearly demonstrated in \cite{GF:70}. Elliptic
functions were successfully employed for a rectangular lattice of circular inclusions \cite{BK:01} as well as 
in the problem of periodic fibrous composites in applications to biological tissues \cite{BC:08}. 
A method of functional equations \cite{Mi:97}, \cite{Ryl:00} employing analytic functions was used to find 
an expression of the permittivity tensor for small volume fraction of inclusions. 

Another method was introduced in \cite{B:78,B:80,B:81,B:82} and is based on the study of the analytic properties of the
effective parameters. This approach was extended in \cite{Mil:80, Mil:81a, Mil:81b}  and proved to be efficient 
for obtaining bounds 
on complex effective parameters. Its mathematical justification is given in \cite{GP:83, GP:85}.

In this paper we represent the potential in terms of Weierstrass' $\zeta$-functions
and their derivatives (an analog of periodically distributed multipoles). This ensures periodicity
of the electric field in the whole plane and avoids the problem of summation of conditionally convergent 
series. Then we determine the average electric field and electric displacement within the parallelogram of 
the periods. It allows us to find an explicit formula for the tensor of effective properties.

\section{Representation of solution  and compliance  with the boundary conditions}

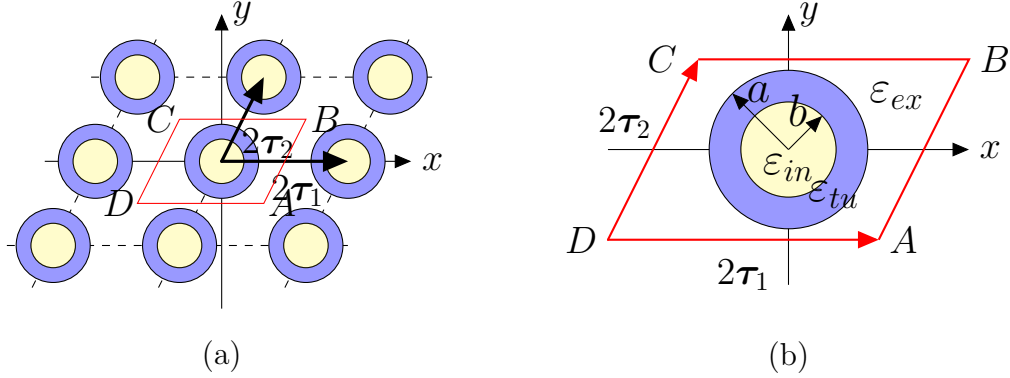
\begin{figure}[H]
  \centering
\begin{minipage}{.3\textwidth}
\hspace*{-15mm}
  \begin{tikzpicture}[>=triangle 45,scale=0.56]

\draw [->] (-4.2,0) -- (4.5,0)  node[right] {\large $x$};
\draw [->] (0,-3.5) -- (0,3.5)  node[right] {\large $y$};

\draw [dashed] (-4.5,-3) -- (-1.5,3);
\draw [dashed] (-1.5,-3) -- (1.5,3);
\draw [dashed] (1.5,-3) -- (4.5,3);

\draw [dashed] (-3.1,2) -- (5,2);
\draw [dashed] (-5.1,-2) -- (3,-2);

\foreach \x in {-3,0,3}
        \draw[fill=blue!40, opacity=1.0,thin] (\x,0)  circle (25pt);
\foreach \x in {-3,0,3}
         \draw[fill=blue!40, opacity=1.0,thin] (\x +1,2)  circle (25pt);
\foreach \x in {-3,0,3}
         \draw[fill=blue!40, opacity=1.0,thin] (\x -1,-2)  circle (25pt);

\foreach \x in {-3,0,3}
        \draw[fill=white!90, opacity=1.0,thin] (\x,0)  circle (15pt);
\foreach \x in {-3,0,3}
        \draw[fill=yellow, opacity=0.2,thin] (\x,0)  circle (15pt);
\foreach \x in {-3,0,3}
         \draw[fill=white!90, opacity=1.0,thin] (\x +1,2)  circle (15pt);
\foreach \x in {-3,0,3}
        \draw[fill=yellow, opacity=0.2,thin] (\x +1,2)  circle (15pt);
\foreach \x in {-3,0,3}
         \draw[fill=white!90, opacity=1.0,thin] (\x -1,-2)  circle (15pt);
\foreach \x in {-3,0,3}
        \draw[fill=yellow, opacity=0.2,thin] (\x -1,-2)  circle (15pt);

\draw [->,very thick,cap=round] (0,0) -- (1,2);
\node [below] at (1.1,1.05) {\large $2\bm{\tau}_2$};
\draw [->,very thick,cap=round] (0,0) -- (3,0);
\node [below] at (1.8,-0.09) {\large $2\bm{\tau}_1$};

\draw [red] ((-1,1) --(2,1) -- (1,-1) -- (-2,-1) -- (-1,1);

\node [right] at (1,-1) {\large $\!A$};
\node [right] at (2,1) {\large $\!B$};
\node  at (-1,1) {\large $C\quad$};
\node  at (-2,-1) {\large $D\quad$};

\node [below] at (0,-4) {(a)};

\end{tikzpicture}
\end{minipage}
\hspace*{10mm}
\begin{minipage}{.3\textwidth}
\begin{tikzpicture}[>=triangle 45,scale=1.2]

\draw [->] (-2,0) -- (2,0)  node[right] {\large $x$};
\draw [->] (0,-1.5) -- (0,1.5)  node[right] {\large $y$};

\draw[fill=blue!40, opacity=1.0,thin] (0,0)  circle (25pt);

\draw[fill=white!90, opacity=1.0,thin] (0,0)  circle (15pt);
\draw[fill=yellow, opacity=0.2,thin] (0,0)  circle (15pt);

\draw [->, thick,cap=round, red] (-2,-1) -- (-1,1);
\node [left] at (-1.4,0.3) {\large $2\bm{\tau}_2$};
\draw [->, thick,cap=round, red] (-2,-1) -- (1,-1);
\node [below] at (-0.5,-1.1) {\large $2\bm{\tau}_1$};

\draw [red, thick] ((-1,1) --(2,1) -- (1,-1); 

\node [right] at (1,-1) {\large $A$};
\node [right] at (2,1) {\large $B$};
\node [left] at (-1,1) {\large $C\;$};
\node [left] at (-2,-1) {\large $D$};
\node at (0.5,-0.5) {\Large $\ei$};
\node at (0.0,-0.2) {\Large $\ec$};
\node at (1.2,0.6) {\Large $\ee$};

\draw [->] (0,0) -- (45:15pt);  
\draw [->] (0,0) -- (135:25pt);  
\node[above] at (55:5pt) {\Large $b$};
\node[above] at (129:15pt) {\Large $a$};

\node [below] at (0,-2) {(b)};

\end{tikzpicture}
\end{minipage}
\caption{(a) A fragment of an infinite periodic array of tubes with the periods $2\bm{\tau}_1$ and $2\bm{\tau}_2$ and 
a fundamental period parallelogram $ABCD$. (b) Material and geometric  parameters  of the tubes. }
\label{fig:array}
\end{figure}

We consider an infinite periodic array of parallel tubes with the periods $2\bt_1$ and $2\bt_2$ (see Figure 
\ref{fig:array}) embedded in a homogeneous medium with dielectric constant $\ee$. Dielectric constant of the tubes
of inner radii $b$ and outer radii $a$ is denoted by $\ei$. We also suppose that the tubes are filled
with a material with dielectric constant $\ec$.
A homogeneous electric field $\E$ is applied in the direction perpendicular to the axes of the tubes.
In the plane of complex variable $z=x+iy$ 
we  introduce the electric potential $u(z)$ which satisfies the equation

\begin{align}
 \nabla \cdot \left[ \ep \nabla u \right] =0, \quad \ep = \left\{
 \begin{array}{l}
  \ec,\quad 0 \leq r < b, \\[2mm]
  \ei, \quad b < r < a, \\[2mm]
  \ee,\quad r >a.
 \end{array}
\right.
\label{main}
\end{align}
On the boundaries $r=a$ and $r=b$ of the tubes we impose continuity conditions
\begin{gather}
  \left\llbracket u \right\rrbracket =0, \label{bc1} \\[2mm]
  \left\llbracket \ep\, \frac{\de u}{\de n} \right\rrbracket = 0, \label{bc2}
\end{gather}
where brackets $\llbracket \cdot \rrbracket$ denote the jump 
of the enclosed quantity across the interface. In addition, we require the field $\nabla u$ to be periodic
\begin{equation}
 \nabla u (z + 2\tau_i) = \nabla u (z), \quad i=1,2,
 \label{per}
\end{equation}
and normalized in such a way that when the radius of the tubes approaches zero the field tends to the homogeneous 
one of intensity $E=E_x -iE_y$
\begin{equation}
  \ue (z) \to -Ez \quad \text {as} \quad a \to 0.
\end{equation}

Following \cite{G:13}, we represent complex potential $u(z)$ in the form
\begin{align}
 \ui (z) &= Ea \sum_{n=0}^\infty \left[ A_n \left( \frac{z}{b} \right)^{2n+1} + B_n \left( \frac{\bar{z}}{b} \right)^{2n+1} \right], \\[2mm]
 \ut (z) &= Ea \sum_{n=0}^\infty \left[ C_n \left( \frac{z}{b} \right)^{2n+1} + D_n \left( \frac{\bar{z}}{b} \right)^{2n+1}
 +  E_n \left( \frac{a}{z} \right)^{2n+1} + F_n \left( \frac{a}{\bar{z}} \right)^{2n+1} \right], \\[2mm]
 \ue (z) &= -Ez + Ea\sum_{n=0}^\infty \frac{a^{2n+1}}{(2n)!} \left[ G_n \zeta^{(2n)} (z) + H_n \zeta^{(2n)} (\bar{z}) \right],
 \label{ue}
\end{align}
where $A_n, \ldots,H_n$ are unknown complex dimensionless coefficients, $\bar{z}$ stands for the complex 
conjugation, and $\zeta^{(2n)} (z)$ is $2n$-th derivative of the Weierstrass $\zeta$-function \cite{Hancock}
\begin{equation}
\zeta (z) = \frac{1}{z} + {\sum_{m,n}}^{\prime} \left[ \frac{1}{z-\P} + \frac{1}{\P}
+ \frac{z}{\P^{\,2}} \right].
\label{zeta} 
\end{equation}
Here  $\P = 2m\tau_1 + 2n\tau_2$.
Prime in the sum means that summation is extended over all pairs $m,\,n$ except $m=n=0$. 
Since the electric field $E$ is periodic, the potential $u(z)$ should be represented as the sum
of periodic and linear functions. The Weierstrass $\zeta$-function has just that property \cite{Cop:48}
\begin{equation}
 \zeta(z+2\tau_k) = \zeta(z) + 2\eta_k, \quad \eta_k = \zeta(\tau_k), \quad k=1,2,
 \label{quasi}
\end{equation}
where constants $\eta_1$ and $\eta_2$ are related by the Legendre identity
\begin{equation}
 \eta_1 \tau_2 - \eta_2 \tau_1 = \frac{\pi i}{2}.
 \label{la}
\end{equation}
Its derivatives however are periodic functions, so that condition \rf{per} is fulfilled.
Also, from \rf{zeta} it follows that
\begin{equation}
  \oint_{ABCD} \zeta^{(n)}(z)\,dz = 0, \quad n \geq 1.
  \label{oint}
\end{equation}

To satisfy conditions \rf{bc1}-\rf{bc2} on the boundary $r=a$ we expand $\zeta(z)$ and its even derivatives
in a Laurent series
\begin{align} 
 \zeta^{(2n)}(z) &= \frac{(2n)!}{z^{2n+1}}- \sum_{k=0}^\infty s_{n+k+1}\,
 \frac{(2n+2k+1)!}{(2k+1)!}\,z^{2k+1},
  \quad n \geq 0, \quad s_1=0,
\end{align}
where 
\begin{equation}
 s_k = {\sum_{n,m}}^\prime \frac{1}{\P^{\,2k}}, \quad k = 2,3, \ldots.
 \label{sk}
\end{equation}
Due to the symmetry of the lattice the only nonzero sums \rf{sk} are those with even powers of $\P$.

Compliance with the boundary conditions \rf{bc1}-\rf{bc2} leads to an infinite system of linear equations
\begin{align}
\label{H}
 H_n &-\gamma_n \sum_{k=0}^\infty
 s_{n+k+1} \frac{(2n+2k+1)!}{(2k)! (2n+1)!}\, G_k\, a^{2n+2k+2} = \gamma_n \delta_{n,0}, \\[2mm]
 \label{G}
 G_n &-\gamma_n \sum_{k=0}^\infty
 s_{n+k+1} \frac{(2n+2k+1)!}{(2k)! (2n+1)!}\, H_k\, a^{2n+2k+2} = 0,
\end{align}
where 
\begin{align}
\label{gamman}
 \gamma_n &= \frac{\alpha -\alphat \nu^{4n+2}}{1 - \alpha \alphat \nu^{4n+2}}, \\[2mm]
\label{alpha}
 \alpha &= \frac{\ei -\ee}{\ei + \ee}, \\[2mm]
\label{alphat}
 \alphat &= \frac{\ei -\ec}{\ei + \ec}, \\[2mm]
 \nu &= \frac{b}{a}, \quad 0 \leq \nu < 1,
 \label{nu}
\end{align}
and $\delta_{n,0}$ is the Kronecker delta.
The other coefficients are expressed through $H_n$ and $G_n$ as follows:
\begin{alignat}{2}
\label{An}
A_n &= \frac{(\alpha -1)(1+\alphat) \nu^{2n+1}}{\alpha -\alphat \nu^{4n+2}}\, H_n, & \quad
B_n &= \frac{(\alpha -1)(1+\alphat) \nu^{2n+1}}{\alpha -\alphat \nu^{4n+2}}\, G_n, \\[2mm]
C_n &= \frac{(\alpha -1)\nu^{2n+1}}{\alpha -\alphat \nu^{4n+2}}\, H_n, & \quad
D_n &= \frac{(\alpha -1)\nu^{2n+1}}{\alpha -\alphat \nu^{4n+2}}\, G_n, \\[2mm]
F_n &= \frac{\alphat (\alpha -1) \nu^{4n+2}}{\alpha -\alphat \nu^{4n+2}}\, H_n, &\quad
E_n &= \frac{\alphat (\alpha -1) \nu^{4n+2}}{\alpha -\alphat \nu^{4n+2}}\, G_n. 
\label{Fn}
\end{alignat}

We introduce new variables
\begin{align}
\label{xn}
 x_n &= H_n + G_n, \\
 y_n &= H_n - G_n.
\label{yn}
\end{align}
Then equations \rf{H}-\rf{G} become independent
\begin{align}
\label{x}
 x_n &-\gamma_n\sum_{k=0}^\infty
 s_{n+k+1} \frac{(2n+2k+1)!}{(2k)! (2n+1)!}\, x_k\, a^{2n+2k+2} = \gamma_0 \delta_{n,0}, \\[2mm]
 \label{y}
 y_n &+\gamma_n\sum_{k=0}^\infty
 s_{n+k+1} \frac{(2n+2k+1)!}{(2k)! (2n+1)!}\, y_k\, a^{2n+2k+2} = \gamma_0\delta_{n,0}.
\end{align}
We will analyze \rf{x}-\rf{y} by the approach described in \cite{G:13}. First, we introduce parameter $h$
\begin{equation}
 h = \frac{a}{\ell}, \quad h \leq \frac{1}{2},
\end{equation}
where $\ell$ is the least distance between the centers of the tubes
\begin{equation}
 \ell = \min (2|\tau_1|, 2|\tau_2|,2|\tau_1 -\tau_2|).
\end{equation}
Then we denote by $S_k$ the dimensionless lattice sums
\begin{align}
 S_k &= {\sum_{n,m}}^\prime \left(\frac{\ell}{\P}\right)^{2k}, \; k=2,3,\ldots, \; S_1 = 0,
 \label{gk}
\end{align}
and represent both equations \rf{x}-\rf{y} as
\begin{equation}
 \x -\G(h) \x = \y,
 \label{Gh}
\end{equation}
where $\x =(u_0,u_1,\ldots) \in \ell_\infty (\C)$, ${\y} =  \gamma_0\, \delta_{n,0}$, 
and operator $\G(h)$ is defined by
\begin{equation}
 \left(\G(h)\, \x\right)_n = \gamma_n \sum_{k=0}^\infty G_{n,k}\, u_k\, h^{2n+2k+2}, 
\end{equation}
where
\begin{align}
 G_{n,k} &=  \pm \frac{(2n+2k+1)!}{(2k)! (2n+1)!} \,S_{n+k+1}, \quad G_{0,0}=0.
\end{align}

Properties of equation \rf{Gh} describes the following
\begin{theorem}
Equation \rf{Gh} has the following properties:
 \begin{enumerate}
  \item [(a)] 
  For each $0\leq h \leq \frac{1}{2}$ $\G (h)$ is a bounded operator in $l_{\infty} ({\C})$.
  \item [(b)] 
  If $0\leq h < \frac{1}{2}$ then operator $\G (h)$ is compact. 
  \item [(c)]
  The norm of $\G (h)$ is estimated by
  \begin{equation}
   \dst \|\G (h) \|_{\infty} \leq \left|\gamma_0 \right| \,  
  \left(\left(\frac{h}{1-h}\right)^{2}+\left(\frac{h}{1+h}\right)^{2} \right)\sup_{n} |S_n|.
  \label{normG}
  \end{equation}
  \item [(d)]
  If $\|\G (h) \|_{\infty} < 1$ then \rf{Gh}  has a unique solution ${\x}_0 \in c_0 (\C)$. Truncated solution 
 of \rf{Gh} converges exponentially to ${\x}_0$  and can be represented as a convergent power series in $h$.
 \end{enumerate}
\end{theorem}
Proof of the theorem is almost identical to that given in \cite{G:12}.

We will seek for the series solution of \rf{Gh} in the form
\begin{equation}
 u_n = \gamma_0\, \delta_{n,0} +\sum_{m=0}^\infty p_{n,m}h^{2n+2m+2}.
 \label{vn}
\end{equation}
Substitution of \rf{vn} into \rf{Gh} gives a recurrence relation for the coefficients $p_{n,m}$:
\begin{align}
 \label{pn0}
 p_{n,0} &= \gamma_0 \,G_{n,0}, \\[2mm]
 p_{n,k} &= \sum_{m=0}^{\left[ \frac{k-1}{2}\right]} G_{n,m} \,p_{m,k-2m-1},
 \label{pnk}
\end{align}
where $[\nu]$ denotes the integral part of $\nu$.

In the next section it will be shown that the effective properties are determined by only $x_0$ and $y_0$ 
in \rf{x}-\rf{y} which we denote as
\begin{align}
\label{lam}
  x_0 &= \gamma_0 \lambda, \quad y_0 = \gamma_0 \mu.
\end{align}
From \rf{pn0}-\rf{pnk} one can find the series expansion for $\mu$ and $\lambda$. The first few terms of their 
expansion are given by
\begin{align}
 \lambda &= 1 + 3 \gamma_0 \gamma_1 S_2^2 h^8+5 \gamma_0\gamma_2 S_3^2 h^{12} +30 \gamma_0 \gamma_1^2 S_2^2 S_3h^{14} \nonu \\[2mm]
 &+ \left(9\gamma_0^2 \gamma_1^2 S_2^4+7\gamma_0 \gamma_3 S_4^2\right)h^{16} 
 +210\gamma_0 \gamma_1 \gamma_2 S_2  S_3 S_4   h^{18}\nonu \\[2mm]
 &+\left(15 \gamma_1 S_2^2 \left( \gamma_0^2 \gamma_2 S_3^2
 +20\gamma_0 \gamma_1^2 S_3^2 \right)+15\gamma_0^2 \gamma_1 \gamma_2 S_2^2 S_3^2 +9 \gamma_0 \gamma_4 S_5^2\right) h^{20}
 +O\left( h^{22} \right),
 \label{lambda}
\end{align}
\begin{align}
 \mu &= 1 + 3 \gamma_0 \gamma_1 S_2^2 h^8+5 \gamma_0\gamma_2 S_3^2 h^{12} -30 \gamma_0 \gamma_1^2 S_2^2 S_3h^{14} \nonu \\[2mm]
 &+ \left(9\gamma_0^2 \gamma_1^2 S_2^4+7\gamma_0 \gamma_3 S_4^2\right)h^{16} 
 -210\gamma_0 \gamma_1 \gamma_2 S_2  S_3 S_4   h^{18}\nonu \\[2mm]
 &+\left(15 \gamma_1 S_2^2 \left( \gamma_0^2 \gamma_2 S_3^2
 +20\gamma_0 \gamma_1^2 S_3^2 \right)+15\gamma_0^2 \gamma_1 \gamma_2 S_2^2 S_3^2 +9 \gamma_0 \gamma_4 S_5^2\right) h^{20}
 +O\left( h^{22} \right).
 \label{mu}
\end{align}

\section{Determination of the effective permittivity tensor}
\setcounter{equation}{0}

Effective permittivity tensor ${\be}^{\ast}$
relates the average electric displacement $\la {\bm D} \ra$ and the average electric field $\la \E \ra$
\begin{equation}
\la{\D}\ra = {\be}^{\ast} \la{\E}\ra.
\label{e}
\end{equation}
Observe that
\begin{align}
 \la{\E}\ra &= \frac{1}{S} \iint_S \E\,dS 
 =  \frac{1}{S} \iint_{S_{in}} \E_{in}\,dS + \frac{1}{S} \iint_{S_{tu}} \E_{tu}\,dS + \frac{1}{S} \iint_{S_{ex}} \E_{ex}\,dS,
 \label{ea}
\end{align}
while
\begin{align}
 \la{\D}\ra &= \frac{1}{S} \iint_S \D\,dS 
 =  \frac{\ec}{S} \iint_{S_{in}} \E_{in}\,dS + \frac{\ei}{S} \iint_{S_{tu}} \E_{tu}\,dS + \frac{\ee}{S} \iint_{S_{ex}} \E_{ex}\,dS,
 \label{da}
\end{align}
where $S$ is the total area of the parallelogram $ABCD$, $S_{in}$ is the disk of radius $b$, $S_{tu}$ is the annular domain
with $b \leq r \leq a$, and $S_{ex}$ is the part of the parallelogram outside the disk $r \leq a$. Thus, in \rf{ea}-\rf{da} we need
to evaluate three distinct integrals.

Using the mean-value property of harmonic functions in the first integral and relations \rf{An}, \rf{xn}-\rf{yn} we get 
\begin{align}
  \iint_{S_{in}} \E_{in}\,dS &=  \E_{in}(0,0) S_{in}=-E\, ab \left( A_0 +B_0, i(A_0-B_0) \right) \nonu \\[2mm]
  &= -\pi b^2 E \frac{(\alpha -1)(1+\alphat)}{\alpha - \alphat \nu^2}
 \left[
 \begin{array}{cc}
  x_0 \\
  iy_0
 \end{array}
\right].  
\end{align}
Evaluation of the second integral gives
\begin{align}
  &\iint_{S_{tu}} \E_{tu}\,dS =- \iint_{S_{tu}} \left( \frac{\de \ut}{\de x}, \frac{\de \ut}{\de y}\right) dS =
   -E\int_b^a \int_0^{2\pi}\sum_{n=0}^\infty (2n+1) \left(\left[ C_n \frac{1}{\nu} \left(\frac{r}{b}\right)^{2n}
  e^{i2n\phi} \right. \right. \nonu \\[2mm]
 & + \left. D_n \frac{1}{\nu} \left(\frac{r}{b}\right)^{2n} e^{-i2n\phi} -E_n \left(\frac{a}{r}\right)^{2n+2} 
e^{-i(2n+2)\phi} - F_n \left(\frac{a}{r}\right)^{2n+2} e^{i(2n+2)\phi}\right],
\left[iC_n \frac{1}{\nu} \left(\frac{r}{b}\right)^{2n} e^{i2n\phi} \right. \nonu \\[2mm]
& - \left. \left.iD_n \frac{1}{\nu} \left(\frac{r}{b}\right)^{2n} e^{-i2n\phi} -iE_n \left(\frac{a}{r}\right)^{2n+2} 
e^{-i(2n+2)\phi} + iF_n \left(\frac{a}{r}\right)^{2n+2} e^{i(2n+2)\phi}\right]\right)rdr d\phi \nonu \\[2mm]
&=  -\pi a^2 E \left( \frac{1}{\nu} - \nu \right) \left( C_0 +D_0, i(C_0-D_0) \right)
= -\pi a^2 E \,\frac{(\alpha -1)(1-\nu^2)}{\alpha - \alphat \nu^2}
 \left[
 \begin{array}{cc}
  x_0 \\
  iy_0
 \end{array}
\right].  
\end{align}

To evaluate the last integral we change the variables form $x,y$ to $z, \bar{z}$ and apply Green's theorem
in complex form 
\begin{align}
  &\iint_{S_{ex}} \E_{ex}\,dS = - \iint_{S_{ex}} \left( \frac{\de \ue}{\de x}, \frac{\de \ue}{\de y}\right)\,dS
  = - (1,i)\iint_{S_{ex}} \frac{\de \ue}{\de z} \,dS - (1,-i)\iint_{S_{ex}} \frac{\de \ue}{\de \bar{z}} \,dS \nonu \\[2mm]
&=\frac{(1,i)}{2i}\left(\oint_{\Pi} \ue\,d\bar{z} -\oint_{C} \ue\,d\bar{z}\right)
- \frac{(1,-i)}{2i}\left(\oint_{\Pi} \ue\,dz -\oint_{C} \ue\,dz\right),
 \label{Eex}
\end{align}
where $\Pi$ is the perimeter of the parallelogram $ABCD$, while
$C$ is the circle of radius $a$. Observe that $\ue = \ut$ when $r=a$, and the integrals over the circle can be evaluated
directly
\begin{align}
 \oint_{C} \ue\,dz = \oint_{C} \ui\,dz = 2\pi i a^2 E \left(\frac{1}{\nu} D_0 +E_0 \right).
\end{align}
The use of quasiperiodicity of $\zeta$-function \rf{quasi} greatly facilitates evaluation of the integrals 
over the parallelogram $ABCD$ (see Figure \ref{fig:array}(b)). We have
\begin{align}
 &\oint_{\Pi} \zeta^{(2n)}(z)\,dz = \int_A^B + \int_B^C + \int_C^D + \int_D^A 
 = \int_D^C \zeta^{(2n)}(z+2\tau_1)\,dz - \int_D^A \zeta^{(2n)}(z+2\tau_2)\,dz \nonu \\[2mm]
&- \int_D^C \zeta^{(2n)}(z)\,dz + \int_D^A \zeta^{(2n)}(z)\,dz
= \int_D^C \left[ \zeta^{(2n)}(z+2\tau_1) - \zeta^{(2n)}(z) \right]dz \nonu \\[2mm]
& - \int_D^A \left[ \zeta^{(2n)}(z+2\tau_2) - \zeta^{(2n)}(z) \right]dz
= \left(2\eta_1 \int_D^C dz - 2\eta_2 \int_D^A dz \right) \delta_{n,0} \nonu \\[2mm]
& = \left(2\eta_1 2\tau_2 - 2\eta_2 2\tau_1  \right) \delta_{n,0}. 
\end{align}
In the same manner we evaluate similar integrals appearing in \rf{Eex} 
\begin{align}
 &\oint_{\Pi} \zeta^{(2n)}(z)\,d\bar{z} = \left(2\eta_1 2\bar{\tau}_2 - 2\eta_2 2\bar{\tau}_1  \right) \delta_{n,0}, 
\\[2mm]
 &\oint_{\Pi} \zeta^{(2n)}(\bar{z})\,d\bar{z} = \left(2\bar{\eta}_1 2\bar{\tau}_2 - 2\bar{\eta}_2 2\bar{\tau}_1  \right) \delta_{n,0}, 
\\[2mm]
&\oint_{\Pi} \zeta^{(2n)}(\bar{z})\,dz = \left(2\bar{\eta}_1 2\tau_2 - 2\bar{\eta}_2 2\tau_1  \right) \delta_{n,0}.
\end{align}
Here we supposed for simplicity that all lattice sums \rf{gk} are real that is true for rectangular and rhombic lattices.

Combining the three integrals in \rf{ea} and using the Legendre identity \rf{la} we obtain
\begin{equation}
 \la \E \ra = \left( \I - \frac{2a^2 \gamma_0}{S}\, {\bm \Psi} \M \right) \E,
 \label{E}
 \end{equation}
where $\I$ is the identity matrix,
\begin{equation}
 {\bm \Psi} = \left[
 \begin{array}{rr}
  \re \eta_1 \im 2\tau_2 & -\im \eta_1 \im 2\tau_2 \\[2mm]
  -\im \eta_1 \im 2\tau_2 & \pi -\re \eta_1 \im 2\tau_2
 \end{array}
 \right],
 \label{Psi}
\end{equation}
and
\begin{equation}
 \M = \left[
 \begin{array}{cc}
  \lambda & 0 \\
  0 & \mu
 \end{array}
\right], \quad
\E = E \left[
\begin{array}{c}
1 \\
i
\end{array}
\right].
\end{equation}
Here we made use that $\im \tau_1 =0$.

Similar calculations for $\la \D \ra$ in \rf{da} give
\begin{equation}
 \la \D \ra = \ee \left( \I + \frac{2\pi a^2 \gamma_0 }{S}\, \M
 -\frac{2 a^2 \gamma_0}{S}\,  {\bm \Psi} \M  \right) \E.
 \label{D}
\end{equation}
Comparing \rf{E} and \rf{D} with \rf{e} we find the effective dielectric tensor
\begin{equation}
 \be^\ast = \ee \left[ \I + \pi \eta \M
 \left(\I -\eta{\bm \Psi} \M \right)^{-1} \right],
 \label{ep_eff}
\end{equation}
where $\dst \eta = \frac{2a^2\gamma_0}{S}$. 
Note that if $\gamma_0 =0$, that is when
\begin{equation}
 \frac{b^2}{a^2} = \frac{(\ei -\ee)(\ei + \ec)}{(\ei + \ee)(\ei - \ec)}
\end{equation}
the two-dimensional effective medium becomes isotropic with $\be^\ast = \ee \I$ for any geometry 
of the lattice and any concentration of the tubes. 

\section{Maxwell's approximation}
\setcounter{equation}{0}

If  $a \ll \ell$ and the interaction between the tubes is weak one can approximate solution of \rf{x}-\rf{y}
by the their right hand side
\begin{align}
 x_n = y_n = \gamma_0 \delta_{n,0}.
\end{align}
As a result,
\begin{align}
 G_n =0, \quad
 H_n =\gamma_0 \delta_{n,0},
\end{align}
and from \rf{Fn}-\rf{An} one can find expression of the potential
\begin{align}
 \label{ui}
 \ui (z) &= \frac{(\alpha -1)(1+\alphat)}{1-\alpha \alphat \nu^2}\,Ez, \\[2mm]
 \label{ut}
 \ut (z) &= \frac{\alpha -1}{1-\alpha \alphat \nu^2}\left(1 + \frac{\alphat b^2}{|z|^2} \right)Ez, \\[2mm]
 \ue (z) &= -Ez \left(1 -\frac{\alpha -\alphat \nu^2}{1-\alpha\alphat \nu^2} \frac{a^2}{|z|^2} \right).
 \label{uea}
\end{align}
The average electric field $\la \E \ra$ in the medium and that in the core $\la \E_{in} \ra$ and the tubes $\la \E_{tu} \ra$ 
are related by
\begin{equation}
 \la \E \ra =  \nu^2 f \la \E _{in} \ra + (1-\nu^2)f \la \E_{tu} \ra  + (1-f)\la \E_{ex} \ra,
 \label{Ea}
\end{equation}
where $f$ is the volume fraction of solid rods of radius $a$.

Similar relation is valid for the average electric displacement $\la \D \ra$ 
\begin{equation}
 \la \D \ra =  \ec \nu^2 f \la \E _{in} \ra + \ei (1-\nu^2)f \la \E_{tu} \ra  + \ee(1-f)\la \E_{ex} \ra.
 \label{Da}
\end{equation}
From \rf{ui}--\rf{ut} we find
\begin{align}
 \label{Ei}
  \la \E _{in} \ra &= -\frac{(\alpha -1)(1+\alphat)}{1-\alpha \alphat \nu^2}\, \E, \\[2mm]
  \la \E _{tu} \ra &= -\frac{\alpha -1}{1-\alpha \alphat \nu^2}\,\E.
 \end{align}
As for $\la \E _{ex} \ra$ we assume that $\la \E _{ex} \ra = \E$. Then from \rf{Ea} and \rf{Da} we obtain
\begin{align}
 \la \E \ra &= (1-\gamma_0 f)\E, \\[2mm]
 \la \D \ra &= \ee (1+\gamma_0 f)\E.
\end{align}
Comparing the two expressions we arrive at the effective dielectric constant 
\begin{equation}
 \ep^\ast = \ee \,\frac{1+\gamma_0 f}{1-\gamma_0 f},
 \label{max}
\end{equation}
where 
\begin{equation}
 \gamma_0 = \frac{\alpha -\alphat \nu^2}{1-\alpha \alphat \nu^2}. 
\end{equation}
Similar to the lattice case \rf{ep_eff}, $\gamma_0 = 0$ implies $\ep^\ast = \ee$.
 As $\nu \to 0$ (solid rods) the formula becomes regular Maxwell's approximation for the two-dimensional case.
 
\section{Regular lattices}

For regular lattices (square or hexagonal) one can show \cite{GF:70} that $\im \eta_1 =0$ and 
$\dst \re \eta_1 \im 2\tau_2 = \frac{\pi}{2}$, so that $\dst {\bm \Psi} = \frac{\pi}{2}\,\I$ in \rf{Psi}.
As a result, $\lambda= \mu$ in \rf{lam}, and $\be^\ast$ becomes an isotropic tensor $\be^\ast = \ep^\ast \I$ with

\begin{equation}
 \ep^\ast = \ee \,\frac{1+\gamma_0 \lambda f}{1-\gamma_0 \lambda f},
 \label{ep_iso}
\end{equation}
where $f$ is the volume fraction of solid cylinders of radius $a$, while $\lambda$ can be calculated either numerically 
form \rf{x} and \rf{lam} or by the series expansion \rf{lambda}. In the latter case for the square array we obtain the 
following expansion
\begin{equation}
 \lambda = 1 +3\gamma_0  \gamma_1 S_2^2h^{8}+\left(9\gamma_0^2 \gamma_1^2 S_2^4+7\gamma_0 \gamma_3 S_4^2\right)h^{16}
 + O(h^{24}),
 \label{lam_square}
\end{equation}
where
\begin{equation}
 S_2 = {\sum_{n,m}}^\prime \frac{1}{(m+in)^4}\approx 3.15121, 
 \quad S_4 = {\sum_{n,m}}^\prime \frac{1}{(m+in)^8}\approx 4.25577.
\end{equation}

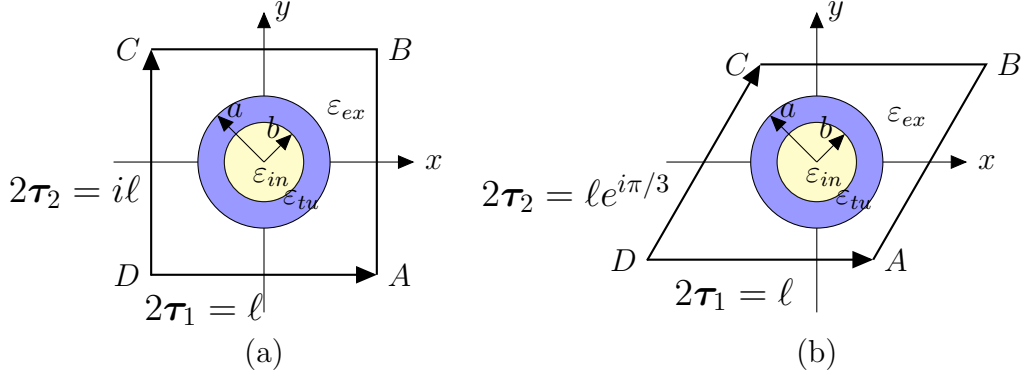
\begin{figure}[H]
  \centering
\begin{minipage}{.3\textwidth}
\hspace*{-25mm}
\begin{tikzpicture}[>=triangle 45,scale=1.0]

\draw [->] (-2,0) -- (2,0)  node[right] { $x$};
\draw [->] (0,-2) -- (0,2)  node[right] { $y$};

\draw[fill=blue!40, opacity=1.0,thin] (0,0)  circle (25pt);

\draw[fill=white!90, opacity=1.0,thin] (0,0)  circle (15pt);
\draw[fill=yellow, opacity=0.2,thin] (0,0)  circle (15pt);

\draw [->,thick,cap=round,black] (-1.5,-1.5) -- (-1.5,1.5);
\node [left] at (-1.5,-0.4) {\large $2\bm{\tau}_2 = i\ell$};
\draw [->,thick,cap=round,black] (-1.5,-1.5) -- (1.5,-1.5);
\node [below] at (-0.8,-1.6) {\large $2\bm{\tau}_1 = \ell$};

\draw [thick] ((-1.5,1.5) --(1.5,1.5) -- (1.5,-1.5); 

\node [right] at (1.5,-1.5) { $A$};
\node [right] at (1.5,1.5) { $B$};
\node [left] at (-1.5,1.5) { $C$};
\node [left] at (-1.5,-1.5) { $D$};
\node at (0.5,-0.5) {$\ei$};
\node at (0.1,-0.2) { $\ec$};
\node at (1.1,0.7) {$\ee$};

\draw [->] (0,0) -- (45:15pt);  
\draw [->] (0,0) -- (135:25pt);  
\node[above] at (55:6pt) {$b$};
\node[above] at (130:17pt) { $a$};

\node [below] at (0,-2.2) {(a)};

\end{tikzpicture}
\end{minipage}
\hspace*{-10mm}
\begin{minipage}{.3\textwidth}
 \begin{tikzpicture}[>=triangle 45,scale=1.0]
\draw [->] (-2,0) -- (2,0)  node[right] { $x$};
\draw [->] (0,-2) -- (0,2)  node[right] { $y$};

\draw[fill=blue!40, opacity=1.0,thin] (0,0)  circle (25pt);

\draw[fill=white!90, opacity=1.0,thin] (0,0)  circle (15pt);
\draw[fill=yellow, opacity=0.2,thin] (0,0)  circle (15pt);

\draw [->,thick,cap=round,black] (210:2.598) -- (120:1.5);
\node [left] at (-1.8,-0.4) {\large $2\bm{\tau}_2=\ell e^{i \pi/3}$};
\draw [->,thick,cap=round,black] (210:2.598) -- (-60:1.5);
\node [below] at (-1.1,-1.4) {\large $2\bm{\tau}_1=\ell$};

\draw [thick] (120:1.5) -- (30:2.598) -- (-60:1.5); 

\node [right] at (-60:1.5) { $A$};
\node [right] at (30:2.598) { $B$};
\node [left] at (120:1.5) { $C$};
\node [left] at (210:2.598) { $D$};
\node at (0.5,-0.5) {$\ei$};
\node at (0.1,-0.2) { $\ec$};
\node at (1.2,0.6) {$\ee$};

\draw [->] (0,0) -- (45:15pt);  
\draw [->] (0,0) -- (135:25pt);  
\node[above] at (55:6pt) {$b$};
\node[above] at (130:17pt) { $a$};

\node [below] at (0,-2.2) {(b)};

\end{tikzpicture}
\end{minipage}
\caption{Cross-sections of elementary cells of the square (a) and hexagonal (b) lattices. In both cases 
$\dst h = \frac{a}{\ell}$.}
\end{figure}

Similar expansion for a hexagonal array gives
\begin{equation}
 \lambda = 1+5\gamma_0 \gamma_2 S_3^2 h^{12} 
     +\gamma_0 \left(25\gamma_0 \gamma_2^2 S_3^4 + 11\gamma_5 S_6^2 \right) h^{24} + O(h^{36}).
 \label{lam_hex}
\end{equation}
Here $\dst S_3 = {\sum_{n,m}}^\prime \frac{1}{(m+ne^{i\pi/3})^6}\approx 5.86303, \;
 S_6 = {\sum_{n,m}}^\prime \frac{1}{(m+ne^{i\pi/3})^{12}}\approx 6.00964$.\\
Comparison of expansions shows that \rf{lam_hex} decays in $h$ faster than \rf{lam_square}. 
Therefore, Maxwell's approximation is more accurate for the hexagonal lattice. It has also been shown 
in \cite{Dehesa:14} that \rf{ep_iso}, when used for the long wave approximation of the effective parameter
of a hexagonal lattice of solid cylinders, is in a very good agreement with numerical calculations.

\begin{figure}[H]
\centering
  \centering
\subfloat[]{\includegraphics[width=0.5\textwidth]{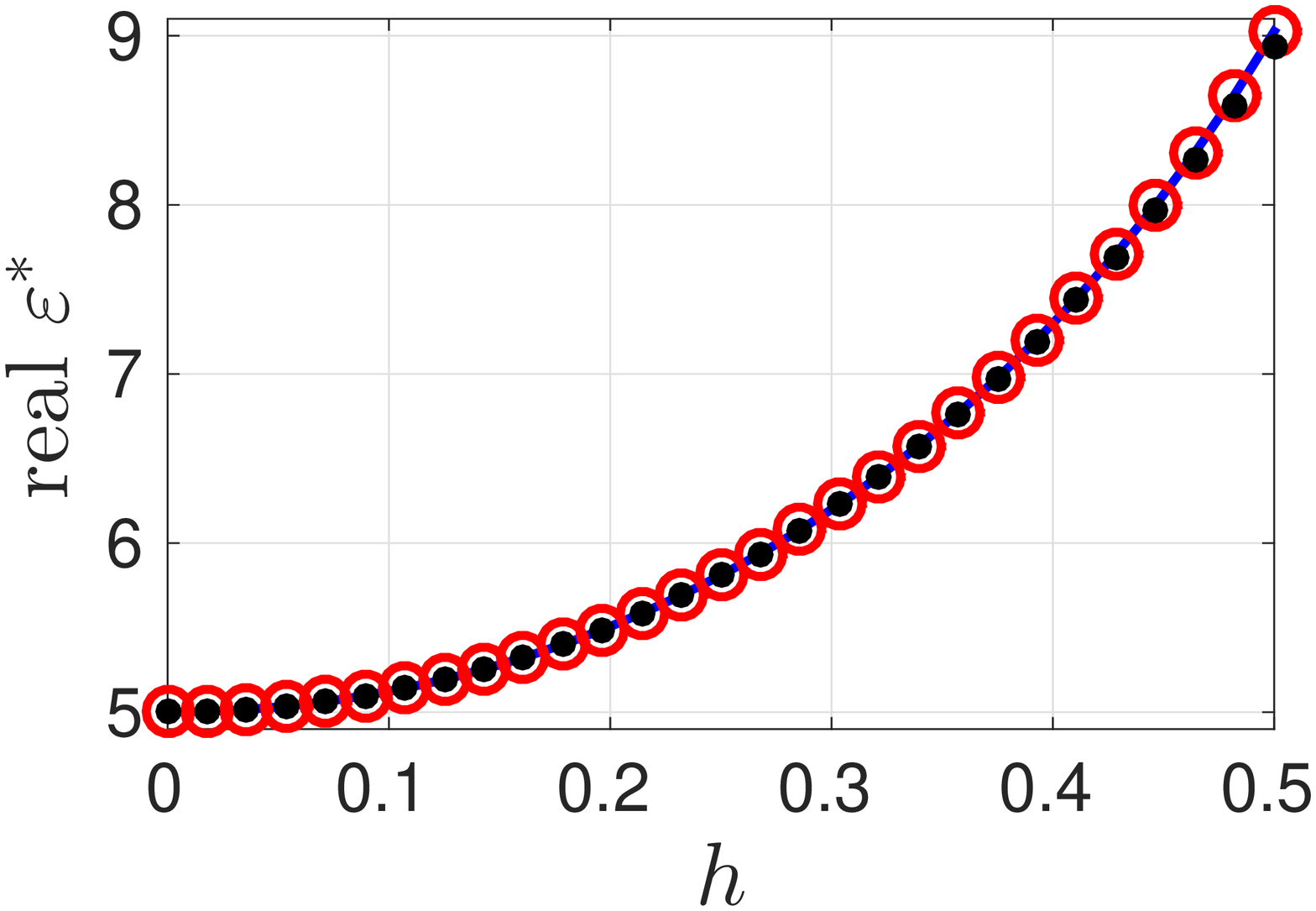}\label{squr_re}}
  \hfill
  \subfloat[]{\includegraphics[width=0.5\textwidth]{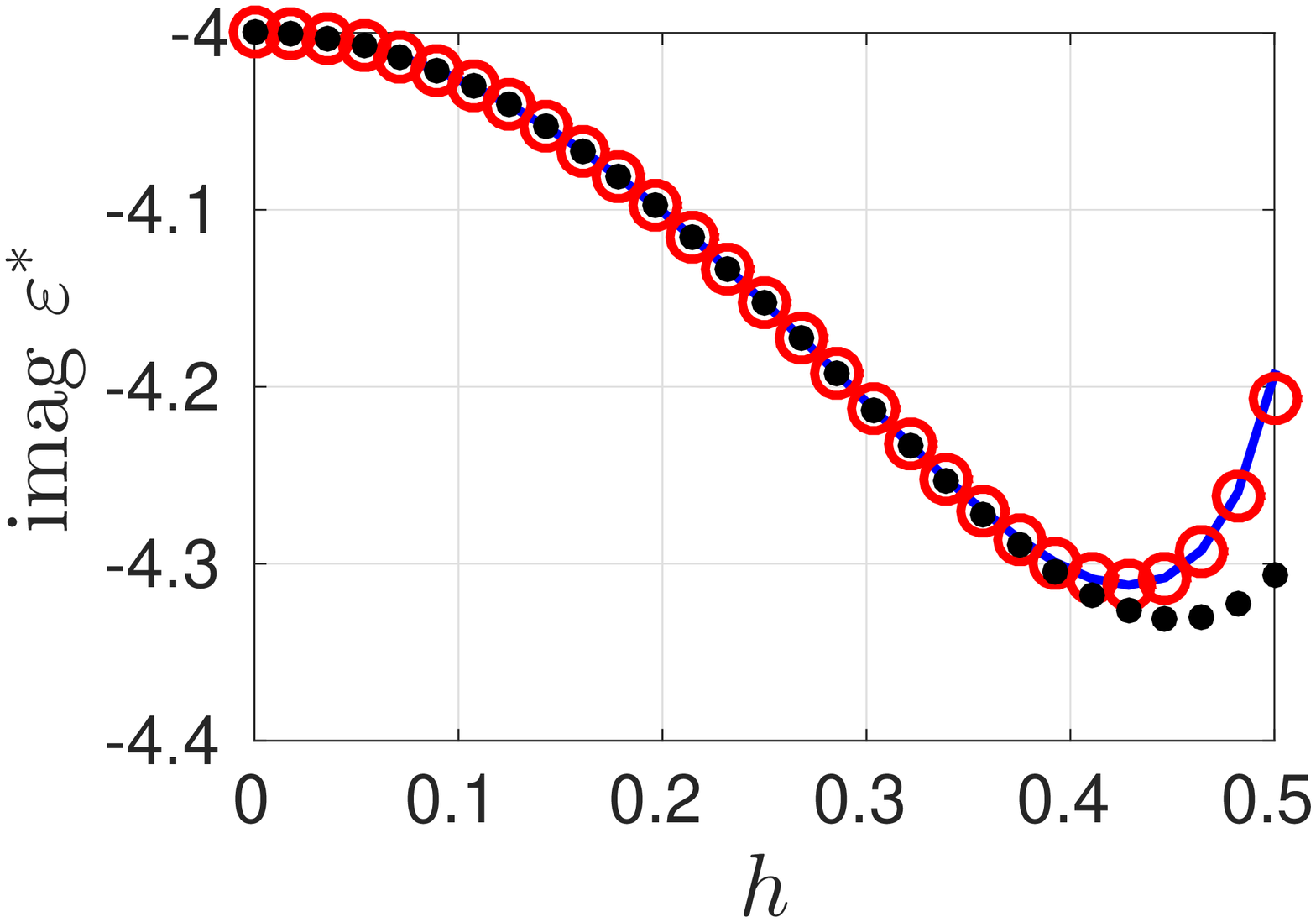}\label{squr_im}}
\caption{Dependence of the real (a) and imaginary (b) parts of the complex effective dielectric constant $\ep^\ast$ of a
square array of tubes on the parameter $\dst h = a/\ell$. The solid blue line corresponds to exact numerical 
evaluation, red circles show result of formulas \rf{ep_iso}--\rf{lam_square}, and black dots represent Maxwell's approximation 
\rf{max} for $\ec = 2-4i$, $\ei = 80-2i$, $\ee = 5-4i$, and $\nu=0.9$.}
\label{fig:square}
\end{figure}

\begin{figure}[H]
  \centering
  \subfloat[]{\includegraphics[width=0.5\textwidth]{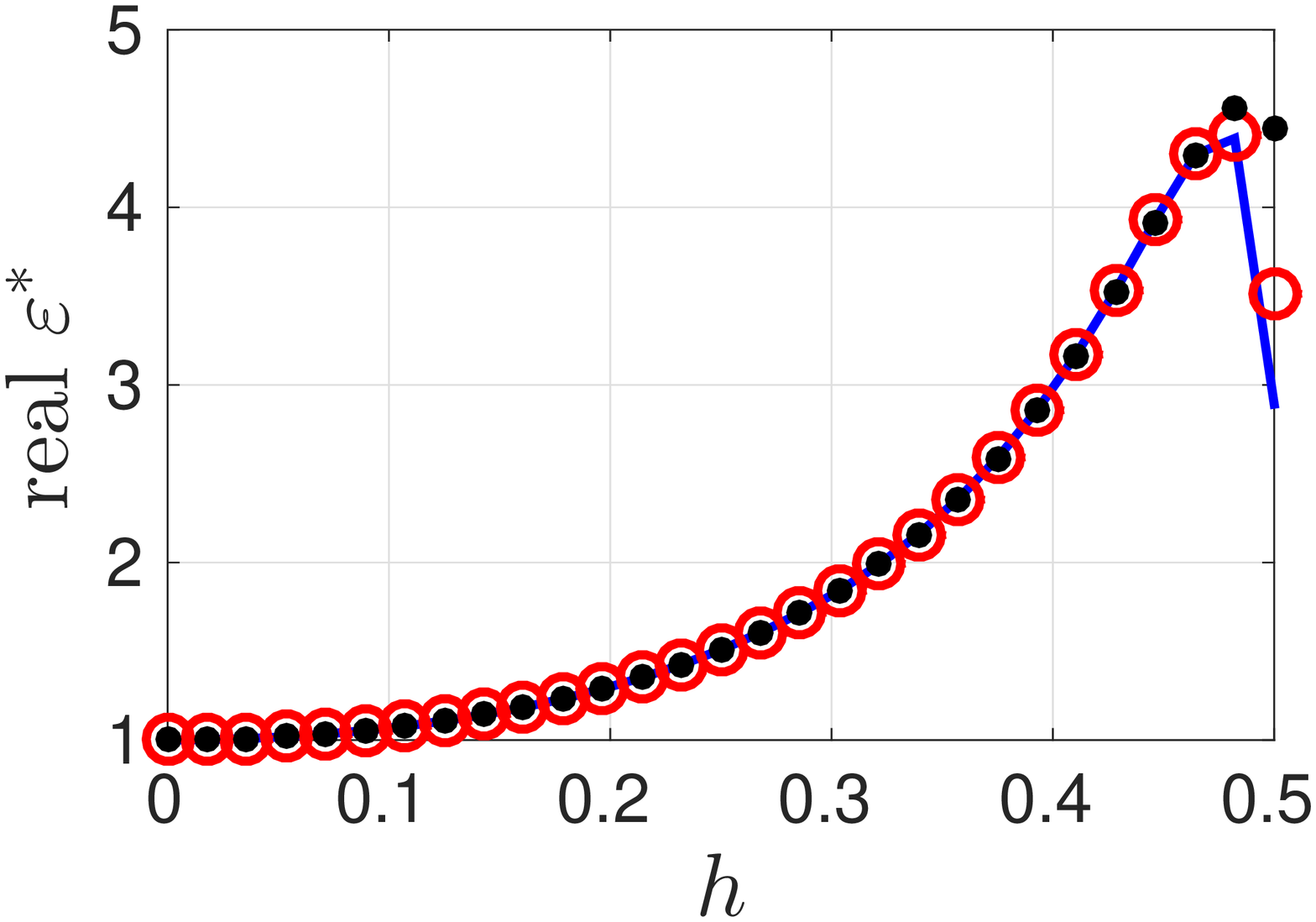}\label{tri_re}}
  \hfill
  \subfloat[]{\includegraphics[width=0.5\textwidth]{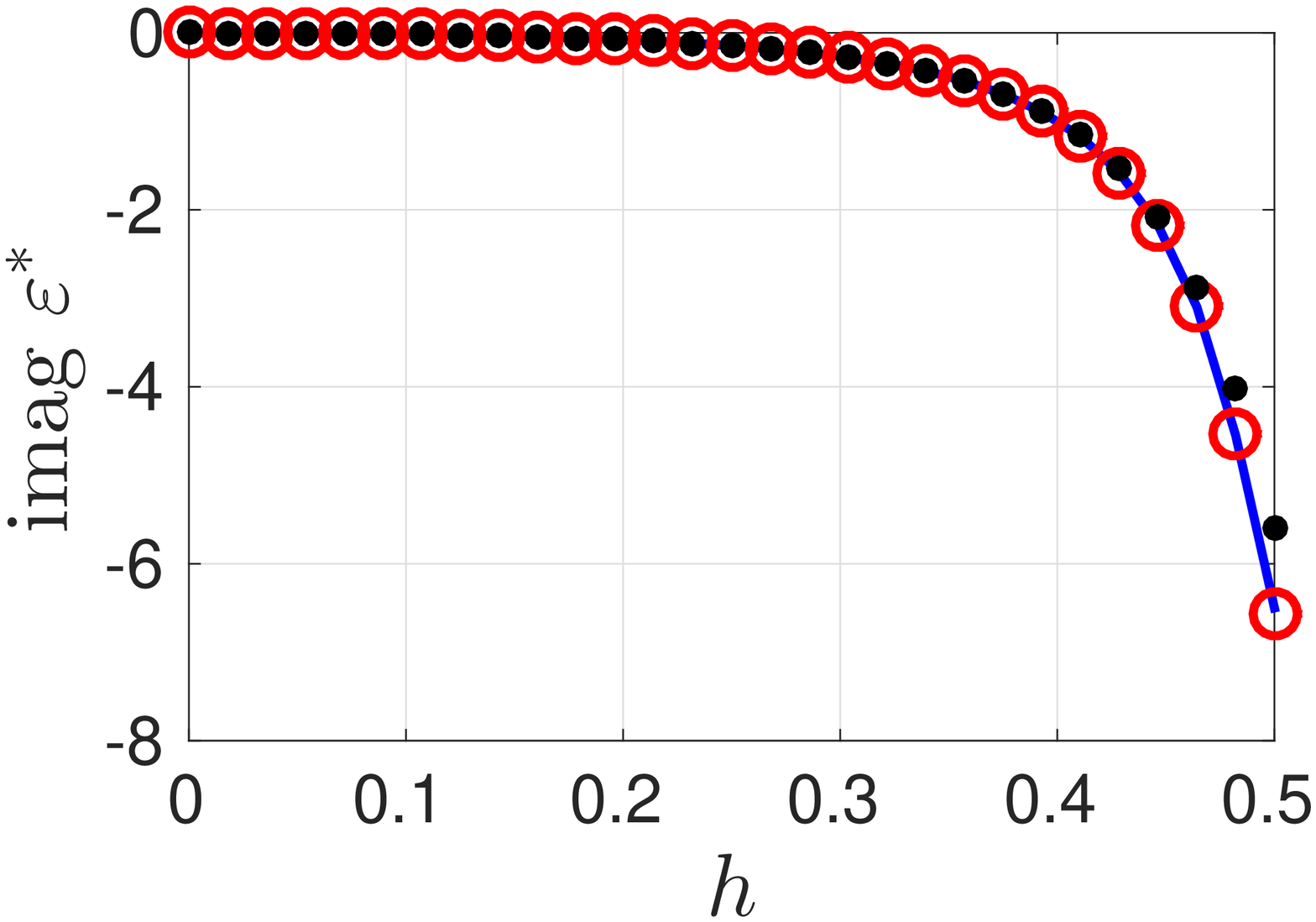}\label{tri_im}}
  \caption{Dependence of the real (a) and imaginary (b) parts of the complex effective dielectric constant $\ep^\ast$ of a
hexagonal array of tubes on the parameter $\dst h = a/\ell$.  The solid blue line corresponds to exact numerical 
evaluation, red circles show result of formulas \rf{ep_iso},\rf{lam_hex}, and black dots represent Maxwell's approximation \rf{max} 
for $\ec = 2-4i$, $\ei = 8-40i$, $\ee = 1$, and $\nu=0.9$.}
\label{fig:triangle}
\end{figure}

Figures \ref{fig:square}-\ref{fig:triangle} show dependence of the real and imaginary parts of the complex effective 
dielectric constant $\ep^\ast$ of a square and hexagonal arrays of tubes on the parameter $\dst h = a/\ell$. 
Formula \rf{ep_iso} gives an excellent agreement between numerical evaluation of $\ep^\ast$ using solution of \rf{x} 
and the expansions \rf{lam_square},\rf{lam_hex} for chosen material parameters. 
In the case of square lattice estimate \rf{normG} gives $\|\G(0.5)\|_\infty \leq 1.4644$ while in fact 
$\|\G(0,5)\|_\infty \approx 0.88035$.
For the hexagonal lattice estimation through \rf{normG} yields $\|\G(0.5)\|_\infty \leq 6.6122$ while direct evaluation
results in $\|\G(0.5)\|_\infty \approx 1.1632$. Maxwell's formula \rf{max} gives a good approximation as long as the norm 
of the operator $\G(h)$ is significantly less than unity. 

\bibliography{tubes}
\bibliographystyle{siam}

\end{document}